\documentclass[useAMS,usegraphicx,usenatbib]{mn2e}

\usepackage{graphicx}
\usepackage{times}
\usepackage{amssymb}
\usepackage{amsmath}
\usepackage{lscape}
\newif\ifAMStwofonts
\AMStwofontstrue

\def\chandra{{\it Chandra}}
\def\hst{{\it HST}}

\def\eclpl{{\rm 3C\thinspace 191}}
\def\eczpa{{\rm 3C\thinspace 294}}
\def\ecaez{{\rm 3C\thinspace 432}}

\def\nir{NIR}

\def\ic{inverse-Compton}
\def\hs{hot-spot}
\def\sync{synchrotron}
\def\xr{X-ray}
\def\B{magnetic}
\def\icb{IC}
\def\ssc{SSC}

\def\frii{\hbox{\rm FR\thinspace II}}

\def\xspecv{{\it XSPEC}{\rm\thinspace v\thinspace 11.3.2}}
\def\ciao{\hbox{\rm CIAO}}
\def\ciaov{\hbox{\rm CIAO\thinspace v3.2.2}}
\def\caldb{\hbox{\rm CALDB\thinspace v3.1.0}}

\def\ks{\hbox{$\rm\thinspace ks$}}

\def\yr{\hbox{$\rm\thinspace yr$}}
\def\myr{\hbox{$\rm\thinspace Myr$}}
\def\hz{\hbox{$\rm\thinspace Hz$}}
\def\mhz{\hbox{$\rm\thinspace MHz$}}
\def\ghz{\hbox{$\rm\thinspace GHz$}}

\def\um{{\hbox{$\rm\thinspace \umu m$}}}

\def\kpc{\hbox{$\rm\thinspace kpc$}}

\def\as{\hbox{$\rm\thinspace arcsec$}}

\def\kmps{\hbox{$\rm\thinspace km~s^{-1}$}}
\def\pcmsq{\hbox{$\rm\thinspace cm^{-2}$}}
\def\kmpspmpc{\hbox{$\rm\thinspace km~s^{-1}~Mpc^{-1}$}}

\def\kev{\hbox{$\rm\thinspace keV$}}

\def\mjypb{\hbox{$\rm\thinspace mJy/beam$}}

\def\jypb{\hbox{$\rm\thinspace Jy/beam$}}
\def\erg{\hbox{$\rm\thinspace erg$}}
\def\ergpcmc{\hbox{$\rm\thinspace erg~cm^{-3}$}}
\def\ergpcmcpkq{\hbox{$\rm\thinspace erg~cm^{-3}~K^{-4}$}}

\def\ergps{\hbox{$\rm\thinspace erg~s^{-1}$}}

\def\msun{\hbox{$\rm\thinspace M_{\odot}$}}

\def\ug{\hbox{$\rm\thinspace \mu G$}}
\def\rpsqm{\hbox{$\rm\thinspace rad~m^{-2}$}}

\def\k{\hbox{\rm\thinspace K}}

\begin{document}

\title[]
{Extended \ic\ emission from distant, powerful radio galaxies}
\author[M. C. Erlund et al.]
{\parbox[]{6.in}
{M.~C. Erlund,$^{1}$\thanks{E-mail: mce@ast.cam.ac.uk} A.~C. Fabian,$^{1}$ Katherine~M. Blundell,$^{2}$ A. Celotti$^{3}$\\ and C.~S.~Crawford$^{1}$}\\\\
\footnotesize
$^{1}$Institute of Astronomy, Madingley Road, Cambridge CB3 0HA\\
$^{2}$University of Oxford, Astrophysics, Denys Wilkinson Building, Keble Road, Oxford OX1 3RH\\
$^{3}$SISSA/ISAS, via Beirut 4, 34014 Trieste, Italy\\
}

\maketitle

\begin{abstract}
  We present \chandra\ observations of two relatively high redshift
  \frii\ radio galaxies, \ecaez\ and \eclpl\ ($z=1.785$ and $z=1.956$
  respectively), both of which show extended \xr\ emission along the
  axis of the radio jet or lobe.  This \xr\ emission is most likely to
  be due to \ic\ scattering of Cosmic Microwave Background (CMB)
  photons. Under this assumption we estimate the minimum energy
  contained in the particles responsible.  This can be extrapolated to
  determine a rough estimate of the total energy. We also present new,
  deep radio observations of \eczpa, which confirm some association
  between radio and \xr\ emission along the NE-SW radio axis and also
  that radio emission is not detected over the rest of the extent of
  the diffuse \xr\ emission. This, together with the offset between the peaks of
  the \xr\ and radio emissions may indicate that the jet axis in this source is
  precessing.

\end{abstract}


\section{Introduction}

Because to its unprecedented sensitivity and spatial resolution, \chandra\
has imaged a wide range of low- and high-redshift radio sources,
revealing their detailed \xr\ emission. Several processes can produce
\xr s over large scales from these objects: thermal emission from
shocks, as well as \sync\ radiation and \ic\ scattering of a seed
photon field (e.g.  \citet{HK02}).  The latter can originate from a
variety of sources.  Synchrotron photons produced in the jet are
up-scattered in the process known as Synchrotron Self-Compton (\ssc).
Thermal photons, reprocessed from the nuclear region, can be scattered
into the jet \citep{scharf4c4117}. A further source of photons is the
Cosmic Microwave Background (CMB), which becomes more important with
redshift \citep{feltenrees}. All such photon fields, once up-scattered
by the relativistic particles in the jet, produce spectra with the
same spectral shape as the synchrotron radiation, because they scatter
off the same population of electrons. If \ic\ is responsible for the
X-ray emission, the radiative lifetimes of (highly energetic) radio
\sync -emitting electrons are typically shorter than for the (less
energetic) electrons which give \ic\ emission in the \xr\ band, so
\ic\ emission traces an older population of particles and may
therefore be offset from, and less spatially compact than, the radio
emission. For \ic\ scattering to dominate over \sync\ emission for a
particular Lorentz factor $\gamma$, the magnetic field energy density
must be less than the energy density of the seed photon field (i.e.
$\frac{B^2}{8\pi}\le\mathcal{U}_{\rm phot}$).  Assuming that this
criterion is met, the photon field with the highest energy density
will be the most important source of \ic\ \xr s.

At redshift $z<1$, the energy density of the CMB is low and \ic\ of
the CMB (\icb) emission is likely to be most important in sources only
if the jets are relativistic (\citealt{0637-752};
\citealt{celottietal2001}). In sources that are not beamed, \icb\ may
be dominated by other \xr\ emission processes \citep{sambrunasurvey}.
\xr\ emission from an increasing number of jets and lobes at these
redshifts has been detected and is often explained by \icb\ (e.g.
\citealt{sambrunasurvey}, \citealt{HK02}, \citealt{elena04},
\citealt{overzier} and \citealt{kands}). Some \xr -bright jets appear
to be closely aligned to the line-of-sight, relativistic and extending
to distances of the order of 100\kpc\ (\citealt{tavecchiosub2kpc} and
\citealt{marshallsurvey}) and appearing, in some cases (e.g. MRC
1136-135, \citet{sambruna2jets}), to be systematically decelerating.

Since the energy density of the CMB increases as ${(1+z)}^{4}$, this
counterbalances surface brightness dimming with redshift. Therefore,
\icb\ in powerful high-redshift objects should be detectable across
the universe, even if beaming is not common \citep{beacons}. So far,
the search for \icb\ in very high-redshift jets has proved elusive:
current snapshot surveys of high-redshift ($z > 3.5$) radio-loud
quasars have generally failed to detect extended \xr\ jet emission
(\citealt{noicz4}, \citealt{laura} and \citealt{beacons}). 

This paper presents \chandra\ observations of two radio galaxies in
the redshift range $1.75-2$ (\ecaez\ and \eclpl), with the aim of
detecting and characterising their \xr\ emission.  These sources, and
\eczpa\ at $z=1.786$ which is re-examined in this paper, are the
highest redshift \frii\ radio galaxies (i.e. \citet{FR} class II
objects) to be detected in \xr s to date\footnote{see
  http://hea-www.harvard.edu/XJET/index.cgi as of 2006 Jan $31$}, with
the exception of 6C\thinspace 0905$+$3955 \citep{6c0905} and
4C\thinspace 41.17 \citep{scharf4c4117}.  New, deep radio observations
of \eczpa\ are also discussed.  The galaxies were chosen because of
their power and spatial extent, meaning that they have a large reservoir of
relativistic particles resolvable by \chandra\ and so are ideal
candidates in the search for \icb. Preliminary reports of these
results appear in \citet{carolinAN} and \citet{xrayuniverse}.  Throughout this
paper, all errors are quoted at $1\sigma$ unless otherwise stated and
the cosmology is $H_{0} = 71$\kmpspmpc, $\Omega_{0}=1$ and
$\Lambda_{0} = 0.73$.


\section{Data Reduction}
\label{sec:reduction}

\begin{table*} 
\begin{tabular}{cccccclclrl}

\hline
Source  & RA (J2000)  & Dec (J2000) & $z$     & \kpc/\as\ & $\rm N_{H,G}$ \pcmsq\ & Date         & Obs ID & (V)FAINT & \ks\  & Reference \\
\hline
\eclpl\ & 08h04m47.9s & +10d15m23s  & $1.956$ & $8.493$ & $2.28\times 10^{20}$ & 2001 Mar. $7$  & 2134 & FAINT  & $8.32$ & \citet{sambrunasurvey}\\ 
        &             &             &         & &                             & 2004 Dec. $12$ & 5626 & VFAINT & $16.65$ & this paper \\
\hline
\eczpa\ & 14h06m44.0s & +34d11m25s  & $1.779$ & $8.544$ & $1.21\times 10^{20}$ & 2000 Oct. $29$ & 1588 & FAINT  & $19.51$ & \citet{3c294one}  \\
        &             &             &         & &                             & 2002 Feb. $25$ & 3445 & VFAINT & $68.49$ & \citet{3c294second}\\
        &             &             &         & &                             & 2002 Feb. $27$ & 3207 & VFAINT & $118.37$ & \citet{3c294second} \\
\hline  
\ecaez\ & 21h22m46.2s & +17d04m38s  & $1.785$ & $8.543$ & $7.40\times 10^{20}$ & 2005 Jan. $7$  & 5624 & VFAINT & $19.78$ & this paper \\
\hline
\end{tabular}
\caption{Source information: name, position, redshift, \kpc\ at redshift of source per\as\ on the sky, Galactic absorption in the direction of the source.  X-ray observation details: date of observation, \chandra\ OBSID, data mode, duration of flare-cleaned observations and references for published data. }
\label{table:obs} 
\end{table*}

Table \ref{table:obs} contains a summary of source information and the
\xr\ observations analysed. The \ciao\ data processing software
package was used for the \xr\ data reduction (\ciaov\ and \caldb).
Pixel randomisation was turned off for all observations using the {\it
  ACIS\_PROCESS\_EVENTS} tool and, when an observation had been taken
in VFAINT mode, {\it check\_vf\_pha=yes} flag was set using the same
tool.  Then the Sub-pixel Resolution Algorithm (\citealt{subpix} and
\citealt{subpix2}) was used to make use of photons that arrive near
the edges and corners of the pixels, as their arrival point can be
determined with sub-pixel resolution.  The algorithm adapts their
positions accordingly.  This improves the half power diameter (HPD) by
approximately ten per cent without losing statistics or affecting
spectral properties. Improving photon positioning improves the image
quality, but not dramatically for these sources. Both \eclpl\ and
\ecaez\ suffer from about 10 per cent pileup in the central source and the
observations were taken in full frame mode with five chips.

In the case of \eclpl, Obs ID $2134$ was reprojected onto the
co-ordinates of $5626$ and for \eczpa, Obs IDs $1588$ and $3207$ were
reprojected onto the co-ordinates of $3445$.

Spectra for the background (an area of sky free from sources near the
target source and on the same chip), the central source, the extended
\xr\ emission and the extended \xr\ emission lying within the radio
contours were extracted separately for each observation and, when
multiple observations were present, stacked using \xspecv.

Despite the limited number of counts, the spectra were fitted with a
Galactic absorbed power-law and, in the case of the nucleus, intrinsic
absorption was also included in the model. $\chi^2$-squared statistics
were used to give a goodness of fit, while C-statistics, which is
appropriate for low photon counts, were used to calculate errors. The
results are presented in Table \ref{table:data}.

The radio observations were made with the VLA\footnote{The National
Radio Astronomy Observatory is operated by Associated Universities,
Inc., under cooperative agreement with the National Science
Foundation.} in A-configuration on the dates listed in Table \ref{table:radio}.  The
data were processed using standard techniques within the AIPS package,
including self-calibration for phase only.

\begin{table} 
\begin{tabular}{ccccl}
\hline  
Source  & frequency  & RMS noise               & linear size & Date                    \\
        & [\ghz]           & [$\times 10^{-4}$\jypb] & [\kpc]      &                        \\
\hline
\eclpl\ & $8.46$           & $1.11$                  & $40$        & 2004 Sept. $18$ \\
\hline
\eczpa\ & $1.43$           & $4.69$                  & $129$       & 2001 Jan. $1$  \\
        & $8.46$           & $0.66$                  & $129$       & 1999 Aug. $13$ \\
\hline
\ecaez\ & $1.54$           & $3.32$                  & $124$       & 1992 Dec. $12$ \\
\hline
\end{tabular}
\caption{All radio data were taken using the VLA in configuration A. RMS is the root mean squared of the background noise in the radio maps.}
\label{table:radio} 
\end{table}


\section{3C\thinspace 191 }
\label{sec:eclpl}

\begin{table*} 
\begin{tabular}{lccccc}
\hline
Target    & photons (bkgrd)              & $\rm {N}_{H}$            & $\Gamma$              & $L_{\rm X}~(2-10\kev)$ & $\chi_\nu^2$ (dof) [ph/bin] \\
          &                              & ${10}^{21}$\pcmsq\       &                       & ${10}^{44}$\ergps\   &                       \\
\hline
\eclpl\   &      ObsID: $2134$ and $5626$    &                          &                       &                      & \\
\hline
nucleus   & $247.96~(0.04);~599.88~(0.12)$ & ${3.8}_{-2.8}^{+2.6}$  & ${1.73}_{-0.12}^{+0.28}$ &  $40.3^{+5.4}_{-0.9}$  & $0.66$ $(44)$ $[15]$ \\
extended  & $21.96~(1.04)~[8]^\star;~60.71~(1.29)~[21]^\star$ & --  & ${1.63}_{-0.35}^{+0.36}$ & $2.33^{+0.65}_{-0.69}$ & $0.49$ $(4)$ $[10]$  \\
          &                              &       --                 & ${1.56}_{-0.31}^{+0.13}$ & $2.40^{+0.54}_{-0.54}$ & C-statistic \\
radio (8.46\ghz)& $11.88~(0.12)~[3]^\star;~27.85~(0.15)~[7]^\star$ & -- & ${1.94}_{-0.21}^{+0.44}$ & $1.24^{+0.40}_{-0.41}$ & C-statistic \\
\hline
\eczpa $^\dagger$  & ObsID: $1588$, $3207$ and $3445$ &                       &              &                          &                      \\
\hline
extended  & $92.73 (10.27);~479.16~(60.84);~302.93~(32.07)$ &  --   & $2.08^{+0.11}_{-0.08}$   & $2.93^{+0.15}_{-0.17}$    & $0.91$~$(60)$~$[20]$ \\
radio ($1.425$\ghz) & $44.88~(3.12);~215.42~(17.58);~136.73~(9.27)$ & -- & $2.00^{+0.12}_{-0.11}$ & $1.41^{+0.01}_{-0.10}$ & $0.54$~$(15)$~$[20]$ \\
radio ($8.46$\ghz) & $14.61~(0.39);~64.67~(2.33);~37.80~(1.20)$ & -- & $1.98^{+0.43}_{-0.39}$ & $0.41^{+0.11}_{-0.11}$ & C-statistic \\
\hline
\ecaez\  &  ObsID: $5624$               &                          &                         &                           &           \\
\hline  
nucleus   & $693.81~(0.19)$              & ${0.21}_{-0.21}^{+0.36}$ & ${1.84}_{-0.11}^{+0.09}$ & $33.81$                 & $0.79$ $(36)$ $[15]$ \\
extended  & $60.63~(1.37)~[24]^\star$    &         --               & ${1.53}_{-0.34}^{+0.35}$ & $1.77^{+0.54}_{-0.47}$  & $0.52$ $(5)$ $[8]$  \\
          &                              &          --              & ${1.57}_{-0.36}^{+0.27}$ & $1.85^{+0.45}_{-0.57}$  &  C-statistic      \\
radio (1.54\ghz) & $35.76~(2.24)~[10]^\star$ &       --             & ${1.52}_{-0.48}^{+0.27}$ & $1.37^{+0.49}_{-0.45}$  &  C-statistic  \\
  \hline
\end{tabular}
\caption{Results of spectral fits. }
\label{table:data} 
\begin{quote}
  \noindent Column 1: region analysed -- extended region includes
  emission inside radio contours.  \\Column 2: contains the background
  subtracted number of source photons in region, in brackets is the
  number of background photons in region.  \\Column 3: the intrinsic
  absorption, which is only relevant to the nucleus of the source.
  \\Column 4: photon index. \\Column 5: \xr\ luminosity in the
  $2-10$\kev\ band rest frame.  \\Column 6: the number not contained
  within brackets is the reduced-$\chi^2$, the number of degrees of
  freedom is in round brackets and the number of photons per bin is in
  square brackets. The $\chi^2$- and C-statistic values are
  consistent. \\ $^\dagger$ A more complete analysis of \eczpa\ can be
  found in \citet{3c294second} and has not been repeated here.
  \\$^\star$ Number of photons due to contamination from the central
  source PSF (i.e. added background).  The $L_{\rm X}$ has been scaled
  accordingly.
\end{quote}
\end{table*}

\begin{figure}
\rotatebox{0}{
\resizebox{!}{8.45cm}
{\includegraphics{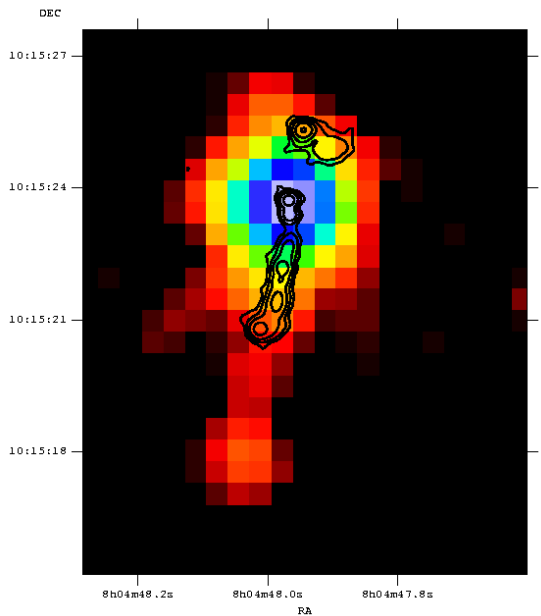}}
}
\caption{Gaussian-smoothed $0.5-3$\kev\ \xr\ image of \eclpl.  The smoothing kernel was 1\thinspace pixel (0.49\as). Overlaid are $8.46$\ghz\ radio contours ($0.4$, $1.3$, $4$, $13$, $40$\mjypb).}
\label{fig:eclpl}
\end{figure}

\eclpl\ is a steep-spectrum, radio-loud, 
non-BAL (Broad Absorption Line) quasar with a central black hole of
$\sim 10^{9}$\msun\ \citep{liu}. High resolution spectroscopy has
shown absorbing gas partially covering the central source: the
absorption lines have flat-bottomed profiles, yet are fully resolved
\citep{hamann}. They are also blue-shifted between $400$ and
$1400$\kmps\ and are consistent with an outflow with a flow-time of
$\sim3\times 10^{7}$ years at a distance of $28$\kpc\ from the quasar
\citep{hamann}.  They seem to arise from a region of the outflow
interacting with the disturbed ISM (Inter-Stellar Medium) in the
elliptical host galaxy \citep{perry}. The optical spectral index,
$\alpha_{\rm opt}=0.7$, indicates that there is little reddening along
our line-of-sight to \eclpl; dust may have been destroyed in shocks
due to the advancing radio source \citep{willott}. \eclpl\ has a
residual rotation measure, which is defined as the rotation measure of 
the source minus the Galactic rotation measure in that direction, 
of $1700$\rpsqm\ \citep{kronberg}. This is two orders of magnitude
larger than other quasars with comparable radio morphology and is
consistent with a thin shell of $\sim 25$\kpc\ across. The magnetic field
parallel to the line-of-sight, calculated from this rotation measure,
is $0.4-4$\ug\ \citep{kronberg}. This source is strongly optically
variable.  It has increased its nuclear \xr\ luminosity by $\sim 20$ per cent 
between the two \chandra\ observations. It is possible that
the nuclear \xr s are relativistically boosted, given the radio
asymmetry and plausible projection.

The extended \xr\ emission is preferentially aligned along the radio
jet direction (Fig. \ref{fig:eclpl}) and, unusually, appears to
continue well beyond the radio emission to the north and south. The
\xr\ emission spans $76$\kpc; almost double that of the radio which is
$40$\kpc. The \xr\ is extended by about $11$\kpc\ to the north-east of
the northern radio \hs: the radio emission goes in the opposite
direction. The extended \xr\ emission to the south is much more
marked: it continues beyond the radio \hs\ by $32$\kpc\ ($3.75$\as).
The photons at the very end of the \xr\ jet seem to be grouped into a
point-like source (hereafter referred to as the tail).  This is not detected by {\it
WAVDETECT} (a wavelet-transform procedure) which only finds the
central source.  In the snapshot Hubble Space Telescope (\hst) data,
there is neither an optical counterpart to the jet
\citep{sambrunasurvey} nor the tail. It is not clear whether the tail
is actually an extension of the jet or a coincidental source.

In order to characterise the tail, the jet is separated into sections:
the extended emission to the north beyond the $3\sigma$ region of the
central source as detected by {\it WAVDETECT}, hereafter referred to
as the northern extension; the photons between the tail and the end of
the $8.46$\ghz\ jet (hereafter referred to as the southern bridge) and
the tail. Considering only observation $5626$ and ignoring
contamination from the central source PSF, the tail (whose centroid
appears to be separated from the rest of the jet by $\sim 2$\as)
contains $6$ photons and has a hardness ratio of
$+0.67^{+0.33}_{-0.49}$. The hardness ratio is a relationship between
the number of photons in the hard ($H$) $2-8$\kev -band and in the soft
($S$) $0.5-2$\kev -band: $\frac{H-S}{H+S}$, calculated using
background subtracted counts.  The southern bridge contains $14$
photons and has a hardness ratio of $-0.76^{+0.32}_{-0.24}$ and the
northern extension has $6$ photons and a hardness ratio of
$-1^{+0.58}_{-0.00}$. There seems to be a gradual change in hardness of
the jet from soft to hard, although the photon statistics are not good
enough to quantify this. Even if the tail were a coincidental source,
the \xr\ jet still extends beyond the $8.46$\ghz\ radio
emission. High resolution MERLIN data at $1.66$\ghz\ follows the
$8$\ghz\ data very closely and is not more extended to the
south.

\xr\ \sync\ emission cannot be ruled out by the extrapolation of the
radio flux in the jet up to \xr\ wavelengths, assuming $\alpha \sim 1$
and an unbroken power-law; however, it is extremely unlikely that a
single unbroken power-law could extend up to \xr\ wavelengths.  The
radiative lifetimes would be extremely short, requiring constant
re-acceleration along the full length of the radio jet. \eclpl\ is not
particularly luminous at IR wavelengths. It has not been detected by
IRAS, and so only an upper limit of its $60$\um\ IR luminosity of
$L_{\rm IR}<4\times10^{46}$\ergps\ is available.  The energy density
of the CMB, which is \mbox{$\mathcal{U}_{\rm CMB}=3.20\times
  10^{-11}$\ergpcmc}, will dominate over the energy density in IR
photons at $\sim 30$\kpc, but it should be remembered that this is an
upper limit. The \xr\ emission aligned along the jet axis is much more
extended and so cannot be explained by up-scattering of nuclear IR
photons. It is also possible for nuclear radiation to be the dominant
\ic\ seed photon field. Both the optical quasar luminosity of $L_{\rm
  nucl}= 2.69\times 10^{46}$\ergps\ \citep{3c191opt}, presumably
produced by accretion, and the luminosity of the Broad Line Region
(BLR), which typically constitutes the dominant scattered field for
relativistic jets on sub-pc scales, $\rm L_{BLR}=1.7\times
10^{46}$\ergps \citep{liu}, are likely to be less important. They
would be dominated by $\mathcal{U}_{\rm CMB}$ at $27$\kpc\ from the
nucleus, which is the size of the extended central source region seen
in Fig. \ref{fig:eclpl} and is at roughly the same distance as the
outflowing shell inferred from other wavelength observations. \xr s
are detected along the full length of the radio jet and beyond, which
implies that they are not generated by \ssc. The extent of the \xr\
emission therefore favours \icb, as does the detection of photons
beyond the end of the jet.

A plausible explanation as to why we detect \xr s beyond the end of
the radio jet is that \eclpl\ could be a double-double radio source
whose jet has been interrupted and restarted \citep{schoenmakers}, possibly by the
same mechanism responsible for the large-scale outflow detected by
\citet{hamann}, which is thought to be $30$\myr\ old.  This is
long enough for the original radio emission (especially at $8.46$\ghz)
to no longer be observable. \icb\ is preferentially emitted from aged
plasma as clearly demonstrated by 6C\thinspace 0905+3955, where \xr s
are visible along the length of the jet emanating from aged plasma, which is
no longer emitting in the radio \citep{6c0905}.  So \icb\ \xr s
provide an excellent tracer of aged synchrotron plasma, and
observations of double-double radio sources would be an interesting demonstration of this with
a combination of \xr\ and radio observations.  It should also be noted
that the $8.46$\ghz\ jet emission is only $40$\kpc -long which is
consistent with it being very young (note that this is the projected length and that this quasar is one-sided).


\section{3C\thinspace 294 }
\label{sec:eczpa}

\begin{figure}
\rotatebox{0}{
\resizebox{!}{8.45cm}
{\includegraphics{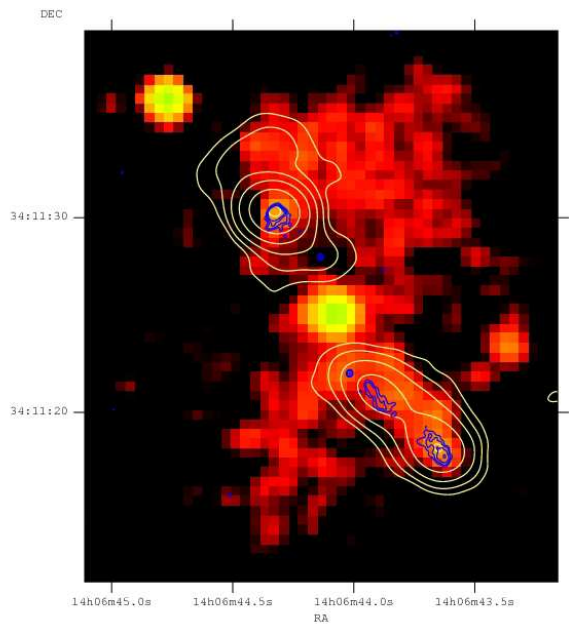}}
}
\caption{Gaussian-smoothed $0.5-3$\kev\ \xr\ image of all three
  observations of \eczpa. The smoothing kernel was 1\thinspace pixel.
  Overlaid with $8.46$\ghz\ radio contours in blue ($0.24$, $1.1$,
  $5.2$, $24$ \mjypb) and yellow $1.425$\ghz\ radio contours ($0.002$,
  $0.006$, $0.02$, $0.06$, $0.2$, $0.6$ \jypb).}
\label{fig:eczpa}
\end{figure}

\begin{figure}
\rotatebox{0}{
\resizebox{!}{8.45cm}
{\includegraphics{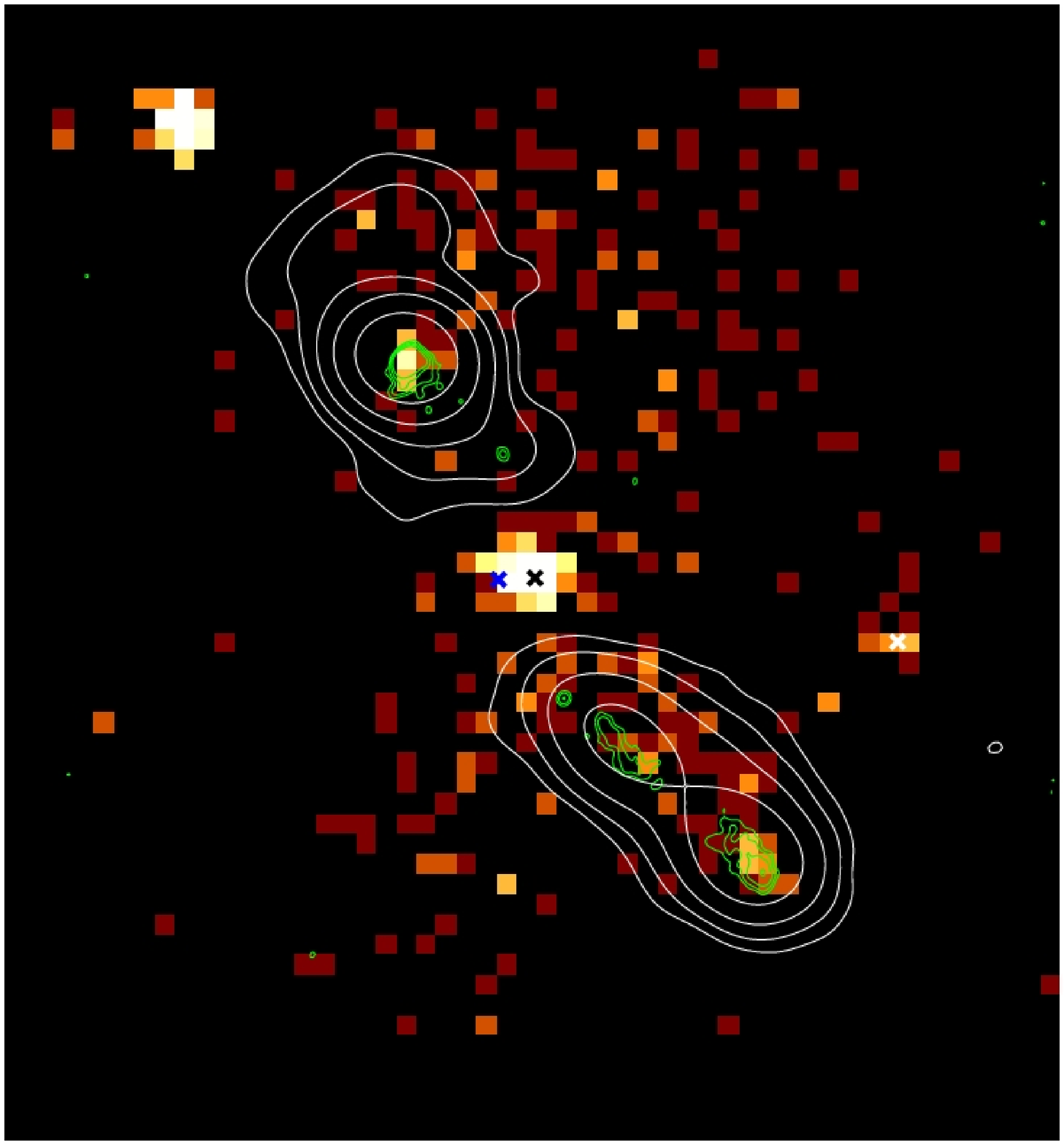}}
}
\caption{$0.5-7$\kev\ \xr\ image of \eczpa. Overlaid with $8.46$\ghz\
  radio contours in green ($0.24$, $1.1$, $5.2$, $24$ \mjypb) and
  white $1.425$\ghz\ radio contours ($0.002$, $0.006$, $0.02$, $0.06$,
  $0.2$, $0.6$ \jypb). The eastern-most cross represents, in dark
  blue, the location of \citet{stockton} eastern stellar object, the
  central cross, in black, is at the position of the radio core and
  western-most cross, in white, represents the position for the AO
  star. All three postitions are taken from \citet{stockton}.}
\label{fig:aspcorr}
\end{figure}

\begin{figure}
\rotatebox{0}{
{\includegraphics[width=0.47\textwidth]{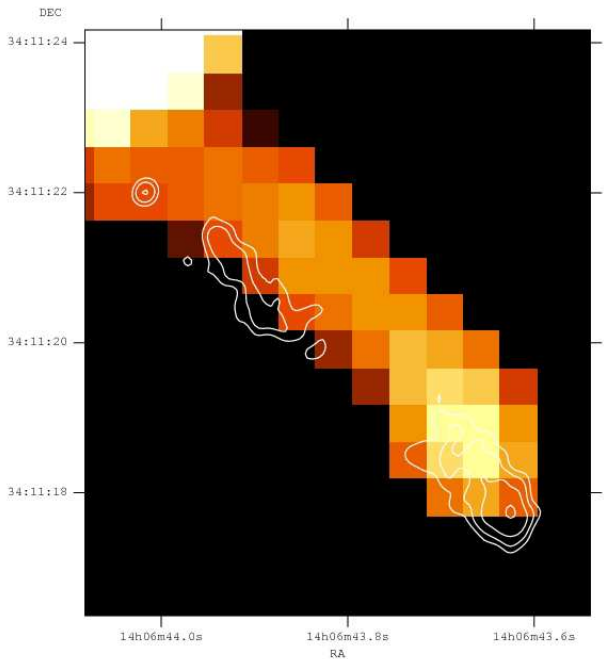}}
}
\caption{Gaussian-smoothed $0.5-3$\kev\ \xr\ image.  The smoothing
kernel was $2$\thinspace pixels.  Overlaid with $8.46$\ghz\ ($0.24$,
$0.40$, $0.67$, $1.1$\mjypb).  The contrast has been adjusted to
highlight the significant offset between the \xr\ and radio jet
emission.}
\label{fig:zoom}
\end{figure}

\begin{figure}
\rotatebox{0}{
{\includegraphics[height=0.47\textwidth]{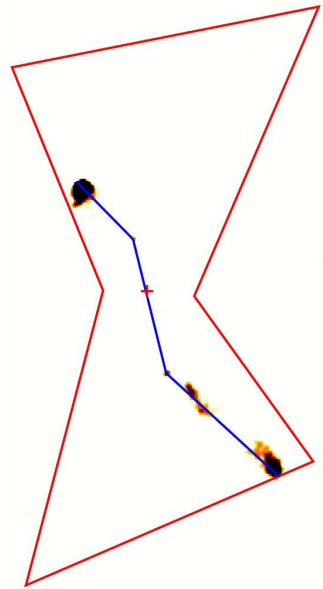}}
}
\caption{This schematic shows the outline of the the extended \xr\
  emission in red, with a red cross marking the \xr\ nucleus.
  Overlaid on the $8$\ghz\ radio image are blue lines showing how the
  jet has changed its orientation.}
\label{fig:schema}
\end{figure}

\eczpa\ is a powerful \frii\ radio galaxy with an intriguingly shaped radio jet (Fig. \ref{fig:schema})
which is highly depolarised, indicating that it lies within a dense
medium. It is embedded in a luminous Lyman-$\alpha$ (Ly$\alpha$) halo,
which has a large velocity shear and is roughly aligned with the radio
source direction. This gas requires a mass equivalent to present day
massive elliptical galaxies and their halos in order for the system to
be gravitationally bound \citep{mccarthy}. An over-density of red
galaxies has also been detected in the near infrared (\nir) in a
$2'\times 2'$ field around \eczpa\ \citep{toft}. \citet{stockton} have
carried out high spatial resolution infrared imaging and find the
central nucleus to have two distinct cores separated by an arcsecond.
Both appear to be Active Galactic Nuclei (AGN); one is compact and
dominated by light from an old stellar population, the other is more
diffuse and contains the principle \xr\ core.  The latter, a partially
obscured AGN, has both a reflection component and a redshifted
6.4\kev\ iron line \citep{3c294second}.  Another AGN, consistent with
being at the same redshift as \eczpa, is $104$\kpc\ to the north-east
of \eczpa.

Bright \xr\ emission associated with the NE-SW radio axis is clearly
detected.  The \xr\ emission is slightly offset and rotated with
respect to the radio emission, as can be seen in Fig. \ref{fig:eczpa}
and Fig. \ref{fig:zoom}.  The \xr\ emission is much more extended than
the radio emission, even in the deeper radio observations presented
here.  In fact, careful inspection of both the $1.425$\ghz\ and
$8.46$\ghz\ radio data shows no radio emission along the NW-SE axis of
the \xr\ source. \citet{3c294second} showed that the surface
brightness profile of this diffuse, hour-glass-shaped emission
declines very steeply at the edges, making it unlikely to be due to
thermal emission: it is best modelled as \icb.

Fig. \ref{fig:schema} shows that the high frequency radio emission to
the north-east and south-west lie parallel to each other. This favours
a change in jet direction due to a change in direction from the
nucleus or a shearing of the radio jet by the ambient medium, rather
than due to a change in ambient pressure \citep{mccarthy}. The
hour-glass \xr\ morphology, together with the current jet axis as seen
in the radio, supports the idea of a precessing jet, such as that
found in the microquasar SS433 \citep{ss433}. This is because the
lifetime of \icb\ electrons is more than an order of magnitude longer
than those responsible for the $8.46$\ghz\ radio emission and,
assuming that the \sync\ number density power-law continues down to
these lower Lorentz factors, there will also be more \icb\ electrons.
The lifetime of the extended \ic\ \xr\ emission suggests a precession
timescale of $>\sim 10^7$\yr. Note that the double nucleus may favour
a precession model \citep{stockton}.

The astrometry of the three observations was corrected using the
Aspect Calculator.\footnote{at
  http://cxc.harvard.edu/cal/ASPECT/fix\_offset/fix\_offset.cgi}
Figure \ref{fig:aspcorr} shows the raw image with crosses
marking the position of central core, the eastern stellar object
(identified by \citet{stockton}) and the position of the AO star (all
positions were taken from \citet{stockton}), all of which match their
\xr\ counterparts well. The northern \xr\ hotspot clearly coincides
with the $8$\ghz\ radio hotspot.  The peak of the \xr\ emission
associated with the southern radio hotspot appears to lie behind the
radio hotspot (see also Fig. \ref{fig:zoom}), but this is within the
errors of \chandra 's $0.6$\as\ positional uncertainty. To the south, the
bulk of the \xr\ emission lies to the north of the $8$\ghz\ emission,
running roughly parallel to the southern radio jet and matching the
radio features.  The offset between the \xr\ and radio jet
is $\sim 1$\as\ (as illustrated in Fig. \ref{fig:zoom}).  This adds
weight to the idea that \eczpa\ is precessing because it appears that
the bulk of the \xr\ emission comes from where the jet was, and not
where it now is.  This is another example of \xr s being emitted from old
plasma as is the case for \eclpl\ and 6C\thinspace 0905+3955.




\section{3C\thinspace 432 }
\label{sec:ecaez}

\begin{figure}
\rotatebox{0}{
\resizebox{!}{6.7cm}
{\includegraphics{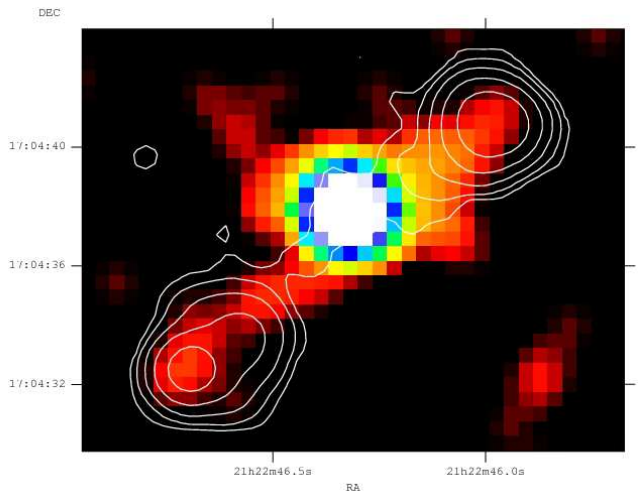}}
}
\caption{Gaussian-smoothed $0.5-3$\kev\ \xr\ image of \ecaez.  The
smoothing kernel was 1.5 pixels ($0.74$\as). Overlaid with $1.54$\ghz\
radio contours ($0.001$, $0.0035$, $0.012$, $0.042$, $0.144$\jypb).}
\label{fig:ecaez}
\end{figure}


\ecaez\ is a powerful, lobe-dominated, \frii\ radio quasar. There is a moderate depolarisation
asymmetry \citep{fernini} demonstrating that, like many \frii\ sources, it is surrounded by a
Faraday-thick magneto-ionic medium (\citealt{laing88}; \citealt{garrington88}).
 The central black hole has a mass of $\sim 5\times 10^{9}$\msun\ as inferred from
Mg\thinspace II line-widths \citep{liu}.

The radio spectral index [$S_\nu \propto \nu^{-\alpha}$] is $\alpha_R
= 0.98$ in the \mhz
-band\footnote{http://www.3crr.dyndns.org/cgi/sourcepage?15},
steepening to $\alpha_R = 1.20$ in the \ghz -band
\citep{richards}. The \xr\ energy spectral index $\alpha_{\rm X}$ is
found from the \xr\ photon index $\Gamma$ in Table \ref{table:data}
using $\alpha=\Gamma-1$; $\alpha_{\rm X}\sim 0.8$ for the nuclear
region and $\alpha_{\rm X}\sim 0.5$ for the extended region.

Fig. \ref{fig:ecaez} shows a Gaussian-smoothed image of \ecaez; apart
from two spurs, one to the east and one to the west of the nucleus
(possibly due to readout streaks from pile-up in the central
source), the \xr\ emission is contained within the $1.54$\ghz\ radio
contours.

\xr\ emission due to \sync\ radiation cannot be ruled out from an
extrapolation of the radio flux, assuming $\alpha \sim 1$ and a
single, unbroken power-law.  The radiative lifetime of these electrons
would be of the order of several hundred years assuming a typical
magnetic field, meaning that even if the electrons were re-accelerated
sufficiently in the \hs, they would not be detectable throughout the
lobe: a lifetime of $> 10^5$\yr\ is necessary for the electrons to
have the time to travel to their detected locations. Synchrotron
emission can be ruled out unless the electrons are being continuously
re-accelerated throughout the whole of the radio lobe: it is hard to
think of a feasible physical theory which would allow this \citep{blundellrawlings2000}. The double-sided 
nature of the \xr\ emission in this source argues against Doppler beaming from a jet.

There are not enough photons to constrain a thermal spectrum, so a
thermal model cannot formally be rejected.  On the assumption that
\ecaez\ is lobe dominated, the \xr s are not produced in shocks along
the jet.  They must be produced by \ic\ up-scattering of a seed photon
field. \ssc\ is important in compact sources, such as at the base of
the jet or the \hs s, so this process is disfavoured by the extended
nature of the \xr\ emission throughout lobe. Another possible seed
photon field is the IR photon field from the nuclear region.  \ecaez\
has not been detected by IRAS so only upper limits on the $60$\um\
luminosity of $L_{\rm IR} < 3\times 10^{46}$\ergps\ exists.
\citep{irasobs}. The energy density of the CMB is
\mbox{$\mathcal{U}_{\rm CMB}=2.52\times 10^{-11}$\ergpcmc} at the
redshift of \ecaez. So the CMB will be the dominant photon field
beyond $\sim 30$\kpc, although this is only an upper limit.  A better
constrained photon field, but one which is likely to be less important
in unbeamed sources such as this one, is the luminosity from the
accretion process in the active nucleus, and that of the BLR. The
latter is estimated at $3.7\times 10^{45}$\ergps\ \citep{liu}, and
thus a typical value for the former would be ten times higher (for a
BLR covering factor $\sim 0.1$). The CMB will dominate over these
photon fields beyond $\sim 35$\kpc\ which means that it will dominate
in the radio lobes in Fig. \ref{fig:ecaez}.  The \xr\ emission extends
well beyond the influence of these two nuclear photon fields: \ecaez\
is $124$\kpc-long. The extent of the \xr s favours \icb\ of the CMB in this
relatively high redshift source.



\section{energy arguments}
\label{sec:energy}
Radio galaxies are thought to trace massive galaxies as they are
often associated with large pools of line-emitting gas comparable to
the envelopes of local cD galaxies \citep{vanojik}.  Some
high-redshift radio galaxies have been associated with proto-clusters
(including \eczpa). Extreme Faraday-rotation measures, similar to those
found in local clusters, along with over-densities of
H$\alpha$-emitters, Ly$\alpha$-emitters and EROs (Extremely Red
Objects), all of which imply a typical cluster environment, are not
confirmed by \xr\ observations. Such systems may not be virialised
structures, so the gaseous environments are not in a deep enough
potential well to emit in the \xr -band. At low redshifts ($z<0.1$),
radio-loud quasars are preferentially found in galaxy groups and
poor-to-moderate clusters \citep{best04}.  Away from the \xr s
associated with lobes and \hs s, soft thermal emission is often
detected with bolometric luminosities of a few times $10^{43}$\ergps,
extending over a few hundred kiloparsecs at the redshift of the quasar
\citep{carolinext}. The influence of radio galaxies on the gaseous
properties of a poor cluster environment would steepen the $L_{\rm
X}-T_{\rm X}$ relationship and so high-redshift powerful radio
galaxies may be in the epoch of pre-heating the proto-cluster
environment. It is therefore important to discover the amount of
energy contained in the jets and lobes of such galaxies.

A lower limit to the amount of energy stored in electrons can be
calculated by assuming that the \xr\ emission detected is indeed due
to \icb.  

The rate at which one electron loses energy in an isotropic radiation
field depends on the energy, $\epsilon_{\rm ph}$, of the photons to be
up-scattered, the Lorentz factor of the electron $\gamma_{\rm e}$
squared, which is roughly the amount by which \ic\ scattering boosts the
initial photon energy, and the rate at which the interaction takes
place. This is given by the Thomson cross-section, $\sigma_{\rm T}$,
times the speed of light, $c$, times the number density of photons available,
$n_{\rm ph}$. $\mathcal{U}_{\rm rad}$ is the energy density of the
photon seed field. The rate of energy loss is therefore given by:
\begin{equation}
  \frac{d\mathcal{E}}{dt} =  n_{\rm ph}\frac{4}{3}\gamma_{\rm e}^2\epsilon_{\rm ph}\sigma_{\rm T} c
                = \mathcal{U}_{\rm rad}\frac{4}{3}\gamma_{\rm e}^2 \sigma_{\rm T} c.
\label{equ:dedt}
\end{equation}

The sources studied here are assumed not to be beamed and to have a
negligible k-correction. On the assumption that CMB photons only
scatter off relativistic electrons with Lorentz factor $\gamma_{\rm e}\sim 10^3$, 
these electrons can be represented as a $\delta$-function rather than the standard power-law distribution.  
This $\delta$-function of electrons, when scattering the
black-body spectrum of CMB photons, produces the $2-10$\kev -band
detected rest-frame \xr\ luminosity.  This is also given by
multiplying Equation (\ref{equ:dedt}) by the number density of electrons,
$N_{\rm e}$ with $\gamma_{\rm e}$
\begin{equation}
  L_{\rm X} = N_{\rm e} \mathcal{U}_{\rm rad}\frac{4}{3}\gamma_{\rm
  e}^{2} \sigma_{\rm T} c.
\label{equ:lx}
\end{equation}
The energy in these relativistic electrons is $\mathcal{E}_{\rm e}$
\begin{equation} 
\mathcal{E}_{\rm e} = N_{\rm e} \gamma_{\rm e} m_{\rm e} c^2,
\label{equ:ee}
\end{equation}
where $m_{\rm e}$ is the rest mass of the electron. Re-arranging Equation (\ref{equ:lx}) 
to give an expression for $N_{\rm e}$ and substituting it into Equation (\ref{equ:ee}) gives
\begin{equation}
  \mathcal{E}_{\rm e} = \frac{3}{4}\frac{L_{\rm X} m_{\rm e} c}{\mathcal{U}_{\rm rad}\gamma_{\rm e}\sigma_{\rm T}} \simeq  \frac{3}{4}\frac{L_{44}}{\gamma_{\rm e}(1+z)^4}10^{64},
\end{equation}
where $L_{\rm X} = L_{44} \times 10^{44}$\ergps\ and $\mathcal{E}_{\rm
e}$ is in \erg.  The results of this calculation for \ecaez,
\eclpl\ and \eczpa\ are presented in Table \ref{table:power}. These
sources have a redshift of $z \sim 2$: the minimum energy in
relativistic particles with $\gamma\sim 10^3$ will therefore be of the
order of $10^{59}$\erg.  This lower limit is important because \icb\
electrons are longer-lived than their radio-\sync -producing
counterparts, giving a better indication of the longevity of the
source and the total energy it injects into its surroundings, which may
have implications for understanding galaxy formation.

The above estimate is strictly a lower limit, as it assumes a
monochromatic (narrow) electron energy distribution.  Assuming that
the electrons responsible for the \xr\ emission are from an electron
population which follows a standard power-law  $N(\gamma) =
{N_{\rm 0}}\gamma^{-p}$, then the energy $E$ originally contained in
these electrons is given by
\begin{equation}
  E = \int^{\gamma_{\rm max}}_{\gamma_{\rm min}}N(\gamma)\gamma m_{\rm
  e} c^2 d\gamma,
\label{equ:eint}
\end{equation}
where $N_{\rm 0}$ is the normalisation of the
power-law. $\gamma_{\rm min}$ and $\gamma_{\rm max}$ are the minimum and maximum 
Lorentz factors assumed to be present in the jet and $p =
2\alpha + 1 = 2\Gamma -1$. Here, $\alpha$ is the \xr\
spectral index and $\Gamma$ is the \xr\ photon index from the \xr\
data fits which are listed, along with $L_{\rm X}$, in Table \ref{table:data}. 

$L_{\rm X}$ is the integrated \ic\ flux, assuming a monochromatic distribution of CMB photons, and can be used to determine $N_{\rm 0}$:
\begin{equation}
L_{\rm X}= \frac{(4/3)^{\alpha}}{2} N_0 \sigma_T c \frac{a
T_{CMB}^4}{\nu_{CMB}^{1-\alpha}}
\frac{\nu^{1-\alpha}_2-\nu^{1-\alpha}_1}{1-\alpha},
\label{equ:lxnew}
\end{equation}
where $h \nu_{CMB} = k T_{CMB}$.
If $\Gamma \sim 2$ (as is the case for \eczpa) then Equation \ref{equ:lxnew} will be replaced by 
\begin{equation*}
L_{\rm X}= \frac{(4/3)^{\alpha}}{2} N_0 \sigma_T c \frac{a
T_{CMB}^4}{\nu_{CMB}^{1-\alpha}} \ln{{\nu_2}\over{\nu_1}},
\end{equation*}
where $\nu_1$ and $\nu_2$ are the frequencies in \hz\ at $2$\kev\ and
$10$\kev\ in the rest-frame respectively. It assumes
that end points of the electron distribution do not contribute and
that the Thomson limit criterion ($\sqrt{\frac{h\nu}{k_{\rm b}T}}\ll
m_{\rm e} c^2$) is upheld, which are both the case over the range in
$\gamma$ of electrons responsible for up-scattering the CMB. $T$ is the temperature of
the CMB at the redshift of the source, given by $T=T_{\rm CMB}(1+z)$,
where $T_{\rm CMB} = 2.728$\k\ is the temperature of the CMB at
$z=0$. $h$ is Planck's constant and $k_{\rm b}$ is the Boltzmann
constant.

Note that this calculation does not require equipartition to
be assumed because the \xr\ emission is used to constrain the
power-law normalisation.  This calculation is, however, highly
sensitive to the value of the \xr\ photon index, $\Gamma$, and its
errors, as well as depending on the choice of $\gamma_{\rm min}$ and
$\gamma_{\rm max}$ which are poorly constrained values.  When
$\gamma_{\rm min}=10^3$, $\gamma_{\rm max}=10^5$ and $\Gamma=1.7$ are chosen, one obtains
$E\sim 1.5\mathcal{E}_{\rm e}$.

\begin{table} 
\begin{tabular}{ccccc}
  \hline  
  Target & radio structure$~^a$ & $\mathcal{E}_e {\rm ~~^b}$ & $B_{\rm max}$  \\
         & [\ghz]              & [$\times 10^{59}$\erg]       & [\ug]          \\
  \hline 
  \ecaez\  & $1.54$            & $1.81\pm 0.93$ & $25$           \\
           & ${\rm ext}$       & $2.44\pm 1.14$ &                \\ 
  \hline
  \eclpl\  & $4.85$            & $1.29.\pm 0.65$ & $28$           \\
           & ${\rm ext}$       & $2.50\pm 1.08$ &                \\
  \hline
  \eczpa\  & $8.46$            & $0.55\pm 0.25$ & $25$        \\
           & $1.425$           & $1.88\pm 0.79$ &                \\
           & ${\rm ext}$       & $3.91\pm 1.42$ &                \\
  \hline 
\end{tabular}
\caption{The C-statistic values of $L_{\rm X}$ in Table \ref{table:data} were used
to calculate $\mathcal{E}_e$.  The errors on $\mathcal{E}_{\rm e}$
were found by sampling a Gaussian distribution about $L_{\rm X}$ with a
standard deviation of the average of the errors on $L_{\rm X}$ and
sampling $\gamma_{\rm e}$ from a log-normal distribution about  $10^3$. $^a$
ext refers to all the extended \xr\ emission, both within and outside
of the radio contours. $^b$ $10^6$ trials were used to calculate the
errors on $\mathcal{E}_{\rm e}$.}\label{table:power} 
\end{table}

So far, only the energy originally in a power-law population of
relativistic electrons has been taken into account.  There is a
considerable amount of debate about the composition of the jet and this
uncertainty can be represented by a factor $k$, $E_{\rm tot} = kE$. 
$k$ can range from $k\sim \mathcal{O}(1)$ for a purely
electron-positron jet, and the energy in the jet is that calculated
above, to $k\sim \mathcal{O}(1000)$ for purely electron-proton jets (for cold protons).
A growing body of work supports the idea of a mainly electron-positron
jet (e.g. \citealt{reynoldscelottirees96} and \citealt{robparticles}),
although, arguments for mainly electron-proton jets can be found in
\citet{celottifabian93} and in \citet{sikora}; the latter work argues
for a proton-dominated jet where the number of electron-positron pairs
greatly exceeds the number of protons in high-power jets typical of
\frii\ sources.  Any entrainment of material surrounding the jet
and/or lobe would also alter the value of $k$ (e.g. for radio sources
in clusters: \citealt{particles} and \citealt{robandygreg}).

If the choice of $\gamma_{\rm min}$ and
$\gamma_{\rm max}$ were correct for the power-law before it was
adiabatically expanded, then the energy in relativistic electrons
calculated here would be an under-estimate; note also that we do not
take into account any $pdV$ work done on any surrounding medium.  An
originally unbroken power-law is assumed and is justifiable because we
would like to know the energy that was originally in the electron population
before any energy losses occurred. This estimate does not take into
account possible re-acceleration, so would be an under-estimate if the
same electrons which, having cooled once, were re-accelerated and
enabled to cool again.

The dominant radiation mechanism, \sync\ or \icb, for the electrons with $\gamma\sim
10^3$ depends on the energy density of the \B\ field,
$\mathcal{U}_{\rm B}$, compared to that of the CMB, $\mathcal{U}_{\rm
CMB}$.  If $\mathcal{U}_{\rm B} \sim \mathcal{U}_{\rm CMB}$, then the
electrons thought to be responsible for up-scattering the CMB would in
fact emit at low \sync\ frequencies.  Therefore, in assuming that electrons with
$\gamma \sim 10^3$ efficiently cool via \icb\ producing the \xr s detected,  
this places an upper-limit on the \B\ field, $B_{\rm max}$, which is given by
\citep{beacons}:
\begin{equation}
B_{\rm max} = (z+1)^2 T^{2}_{\rm CMB} (8\pi {\rm a})^{\frac{1}{2}},
\label{equ:bmax}
\end{equation}
where the radiation density constant is $a=7.5646\times
10^{-15}$\ergpcmcpkq.  This puts a constraint on the magnetic field
energy density which, together with the energy contained in particles,
gives the total energy in the jet \citep{particles}:
\begin{equation}
E_{\rm jet} = kE + Vf\frac{B^2}{8\pi} = kE + E_{\rm B},
\end{equation}
where $f$ is the volume filling fraction and $V$ is the volume of the
lobe. Assuming $f=1$, $V$ is a cylinder on the sky whose diameter and
length are the dimensions of the extracted regions for extended \xr\
emission.  This is an over-estimate of the volume, and thus of the
total energy in the magnetic field. Nonetheless, for both \ecaez\ and
\eclpl, the energy in the particles responsible for the \xr\ emission,
$\mathcal{E}_{\rm e}$, which is a robust lower limit to the energy
stored in relativistic particles in the jet, is roughly a factor of
two greater than the energy in the magnetic field, $E_{\rm B}$. For
\eczpa, where the volume $V$ is represented by two cones, $E_{\rm B}
\sim \mathcal{E}_{\rm e}$.  This means that the energy stored in
relativistic particles $E$ is greater than that stored in the magnetic
field, $E_{\rm B}$. An important caveat is that we assumed that the
emission from these jets is not dominated by relativistic bulk motion;
if it was then the estimated particle number could dramatically
change.


\section{Conclusions}
\label{sec:conclusions}

The \chandra\ observations of two relatively high-redshift \frii\ radio
galaxies, \ecaez\ and \eclpl, show extended \xr\ emission which is
most convincingly explained by inverse-Compton scattering of CMB
photons.  \eclpl\ appears to produce \xr\ emission from beyond the end
of the radio jet, which appears to be harder than the rest of the
extended emission.  The possibility of a coincidental background or
foreground source cannot be excluded.

New radio data for \eczpa\ is presented, confirming the results from
\citet{3c294second} and illustrating that the \xr\ emission is much
more extended than the radio emission. This and the offset detected
between the radio jet axis, which represents current jet activity and
the \xr\ emission, which traces the jet's previous position, provides
some evidence that this source is precessing.

The \xr\ luminosity of the extended regions of these sources is used
to calculate the energy contained in the electrons responsible for the
\xr\ emission. This gives a lower limit to the amount of energy in the
lobe of the order of a few times $10^{59}$\erg\ if \xr\ producing
electrons are assumed to have a Lorentz factor $\gamma \sim 10^3$. The
energy estimate increases by a factor of $\sim 1.5$ when a power-law
distribution of electrons is considered and possibly by up to $10^3$
when the proton component is included. \\

\section*{Acknowledgements}
MCE acknowledges PPARC for financial support. ACF, CSC and KMB thank the Royal
Society. AC acknowledges the Italian MIUR for
financial support. We also thank the referee for their helpful
comments.

\bibliographystyle{mnras} 
\bibliography{mn-jour,Erlundetal05.bib}
\end{document}